# Solid Solution Strengthening and Softening Due to Collective Nanocrystalline Deformation Physics

Timothy J. Rupert[*]

Department of Mechanical and Aerospace Engineering, University of California, 4200 Engineering Gateway, Irvine, California 92697-3975, USA

[*] Corresponding Author. Tel.: 949-824-4937; e-mail: trupert@uci.edu

**Abstract**

Solid solution effects on the strength of the finest nanocrystalline grain sizes are studied with molecular dynamics simulations of different Cu-based alloys. We find evidence of both solid solution strengthening and softening, with trends in strength controlled by how alloying affects the elastic modulus of the material. This behavior is consistent with a shift to collective grain boundary deformation physics, and provides a link between the mechanical behavior of very fine-grained nanocrystalline metals and metallic glasses.

Polycrystalline metals with average grain sizes smaller than 100 nm, commonly referred to as *nanocrystalline*, are promising structural materials due to reports of improved mechanical properties such as strength [1], fatigue resistance [2], and wear resistance [3]. Unfortunately, pure nanocrystalline metals often exhibit limited structural stability, with a number of studies showing evidence of room-temperature [4] and stress-driven grain growth [5], along with a corresponding degradation of strength. To limit this grain growth, processing scientists use the addition of alloying elements to stabilize nanocrystalline microstructures through either kinetic or thermodynamic constraints [6]. Some alloy systems, such as Ni-P [7], rely on complete segregation of the alloying element to grain boundary sites, while others, such as Ni-Fe [8] and Ni-W [9], take advantage of elements which have a subtle tendency to segregate to interfaces. For example, while the grain boundaries in nanocrystalline Ni-W have slightly more W than the grain interior, up to ~20 at.% W can still be incorporated into the Ni lattice [9]. The benefit of subtle grain boundary segregation is that grain size ($d$) becomes a function of dopant concentration, allowing $d$ to be tuned in a controlled manner [10].

Since alloying elements are key ingredients for the production of stable nanocrystalline metals, a detailed understanding of the effect of alloying on mechanical properties is needed. However, the theories which describe solid solution strengthening in coarse-grained alloys, such as those from Fleischer [11] and Labusch [12], are based on the concept that dislocations move freely through the grain interior and that strengthening results from local interactions with solute atoms due to size and elastic modulus mismatches. However, nanocrystalline metals plastically deform through novel physical mechanisms which are dramatically different than those associated with traditional metallic plasticity. For nanocrystalline grain sizes between approximately 15 and 100 nm, plasticity is controlled by nucleation and pinning of dislocations at grain boundary sites [13]. Rupert et al. addressed solid solution effects for these grain sizes by adding a grain boundary pinning term to traditional Fleischer theory, finding that such a model describes many nanocrystalline alloy data sets available in the literature [14]. An interesting corollary of this work was that solid solution softening was predicted for alloy combinations where solute addition either significantly decreases the elastic modulus or the lattice constant of the solvent. However, for grain sizes below ~15 nm, grain boundary sliding and grain rotation become the dominant carriers of plastic deformation [15]. Without appreciable dislocation activity in these materials, no models currently exist to describe how solid solution alloying will affect the strength of the finest nanocrystalline metals.

Probing solid solution effects in extremely fine-grained nanocrystalline alloys is difficult to study experimentally because, as mentioned above, nanocrystalline solid solution alloys often have grain sizes which are intimately tied to composition. In this study, we use molecular dynamics (MD) simulations, where sample composition can be tuned independently of grain size, to systematically explore how alloy chemistry affects the strength of very fine-grained nanocrystalline metals. MD simulations were performed with the Large-scale Atomic/Molecular Massively Parallel Simulator (LAMMPS) code [16] using an integration time step of 2 fs. Nanocrystalline specimens with 100 grains and $d = 5$ nm were created using a Voronoi tessellation method modified to enforce a minimum separation distance between grain nucleation sites. The Cu-Pb and Cu-Ni systems were chosen as model alloys to allow for the systematic variation of elastic modulus and lattice parameter. Pb is more compliant and has a larger lattice spacing than Cu, while Ni is stiffer and has a smaller lattice constant than Cu. Cu-Pb structures were simulated using an embedded atom method (EAM) potential from Hoyt et al. [17], while two sets of Cu-Ni alloys were simulated using EAM potentials from Foiles [18] and Bonny et al.

[19]. The Cu-Ni system is especially interesting, as it exhibits full solid solubility, allowing a large range of alloy compositions to be accessed. All potential files were obtained from the NIST Interatomic Potentials Repository [20].

Randomly selected Cu atoms were replaced with solute atoms to create a solid solution of the desired composition. All simulations employed periodic boundary conditions and a Nose-Hoover thermo/barostat. Each nanocrystalline specimen was equilibrated at 300 K and zero pressure for 100 ps until a steady-state system energy was reached. Figure 1 shows a representative atomic configuration taken from a Cu-4 at.% Pb sample. In this image, Cu atoms in the grain interior, identified by common neighbor analysis (CNA) with an adaptive cut-off value [21, 22], are grey, Cu atoms in the grain boundaries are white, and Pb atoms are red. The Pb atoms are randomly distributed throughout the specimen, with equal concentrations in grain interior and grain boundary regions. While other authors have provided excellent insight into the effect of solutes that segregate to the grain boundaries (see, e.g., [23, 24]), these samples allow us to study random solid solutions.

Uniaxial tensile deformation of each alloy was simulated by applying strain in one direction at a constant true strain rate of $5 \times 10^8$ $s^{-1}$, while keeping zero stress on the other simulation cell axes. Figures 1b and c show a section of a pure Cu sample at 5% tensile strain with the atoms colored according to CNA and von Mises shear strain, respectively. Crystalline atoms are green in Figure 1b. It is clear from a comparison of these two figures that the majority of plastic strain is accommodated at the grain boundaries, although occasionally a stacking fault from partial dislocation propagation can be found. Representative stress-strain curves are shown in Figure 2a and b for selected Cu-Ni and Cu-Pb alloys, respectively. For Cu-Ni alloys, the addition of more Ni leads to a progressive increase in strength. On the other hand, the Cu-Pb alloys exhibit pronounced solid solution softening as Pb content is increased. Such behavior is inconsistent with the models used to describe coarse-grained behavior, which always predict strengthening with solute addition. Yield strength, measured by taking the 1% offset yield stress to allow for the extended microplasticity regime observed by Brandstetter et al. [25], was extracted from each curve and plotted as a function of composition in Figure 3a. Cu-Pb samples with up to 12 at.% Pb were simulated, while Cu-Ni alloys with up to 14 and 100 at.% Ni (i.e., pure Ni) were simulated with the Foiles and Bonny potentials, respectively. All of the data shows that strength changes with composition in an approximately linear fashion, with Cu-Pb strength quickly decreasing and Cu-Ni strength slowly increasing. In contrast, the strength of coarse-grained Cu-Ni increases initially, reaches a peak strength at ~50 at.% Ni, and then decreases towards the strength of pure Ni (i.e., the strength of pure Cu and Ni always increases as solute is added, until an intermediate strength value is reach at an equal mixture of the two elements) [26].

We next quantify changes to the elastic properties and lattice size as the samples are alloyed, since all solid solution theories to date highlight the importance of these properties. Visual inspection of the early, elastic portion of the stress-strain curves in Figure 2 shows that alloying can make the nanocrystalline system either significantly stiffer or more compliant depending on the choice of alloying element. Young's modulus values, extracted from linear fits up to 1% strain, are shown as a function of composition in Figure 3b. Ni stiffens the Cu lattice while Pb makes it more compliant. To observe the effect of alloying on lattice size, we measure the Burgers vector for the Cu-Pb and Cu-Ni (Bonny potential) alloys by measuring the location of the first peak in the radial distribution function. Figure 3c shows that Ni addition decreases the Burgers vector of Cu, while Pb increases it and swells the lattice.

The nanocrystalline pinning model introduced by Rupert et al. [14] for larger nanocrystalline grain sizes places equal emphasis on changes in lattice stiffness and size, with increases in both of these properties leading to higher strengths and decreases leading to softening. If such a model were to describe our Cu-based alloys, one would expect the changes to Young's modulus and Burger's vector to balance each other out and for there to be limited changes in strength as composition is altered. However, our strength data closely follows the trends observed for changes to the elastic modulus, suggesting that elastic properties alone may control strength at these extremely fine nanocrystalline grain sizes. To investigate this more closely, we plot strength as a function of Young's modulus in Figure 4. All of the data from our simulations fall along a straight line with the form:

$$\sigma_y = A \cdot E \quad (1)$$

where $\sigma_y$ is yield strength, $A$ is a fitting constant, and $E$ is Young's modulus. The constant $A$ is equal to 0.0242 here, but we expect it to be a function of applied strain rate and grain size. If Eq. 1 describes the strength of a nanocrystalline alloy, one can then isolate the strengthening/softening increment from solute addition, $\Delta\sigma_{nc,SS}$:

$$\Delta\sigma_{nc,SS} = A \cdot \left(\frac{\partial E}{\partial c}\right) \cdot c \quad (2)$$

Eq. 2 suggests that strength should change with composition, $c$, in a linear fashion and the slope of such a line should only depend on the rate of change of Young's modulus with alloying. To test this hypothesis, we plot Eq. 2 in Figure 3a as dashed lines, after extracting the rate of change of $E$ from Figure 3b, and find a good fit for all three data sets.

To test if Eq. 1 is generally applicable and can be used for other nanocrystalline metals, we plot data from prior MD deformation simulations of pure Ni by Rupert and Schuh [27] in Figure 4 as well. This sample contained only 24 grains and used a different interatomic potential [28], but the grain size ($d = 5$ nm) and applied strain rate ($5 \times 10^8$ s$^{-1}$) were consistent with this study. Although the pure Ni sample has a much higher Young's modulus than the alloys studied here, due to the use of a different EAM potential, this specimen follows the same trend and is described by Eq. 1. Figure 4 tells a consistent story for a range of nanocrystalline metals: elastic stiffness alone controls strength.

Strength that is controlled by the elastic modulus of the material is also found in metallic glasses. Inoue and Takeuchi [29] reviewed the mechanical properties of a variety of amorphous alloys, finding that tensile strength increased linearly with increasing Young's modulus following a form similar to Eq. 1, with a fitting constant of 0.02. This constant is of the same order of magnitude as the fitting constant describing our MD simulations in Figure 4, but the difference between applied strain rates used for these two data sets (quasi-static for the metallic glass literature versus high strain rates from our MD simulations), makes a more detailed comparison difficult. Johnson and Samwer [30] found a similar relationship between the shear strength and the shear modulus of metallic glasses, and these authors also formed a more nuanced theory which incorporated the homologous temperature of each alloy to provide a slightly better description of the strength-modulus relationship. In any case, the strength of both metallic glasses and very fine-grained nanocrystalline metals can be described to first order simply by elastic modulus. Other mechanical properties, such as the pressure sensitivity of strength [31] or the tendency for catastrophic shear banding [32], also suggest a similarity between these two types of materials.

The similarity between the mechanical behavior of nanocrystalline metals and metallic glasses can be understood by comparing their dominant plastic deformation mechanisms. In metallic glasses, there is no long range structural order, so dislocations cannot provide a low-energy pathway for plastic strain. Instead, plastic strain is accommodated by the local shear rearrangement of small groups of atoms [33], in what are called shear transformation zones (STZs). In an STZ, a cluster of atoms cooperatively reorganizes under the action of an applied shear stress, with atoms in one half of the cluster sliding over atoms in the other half. In nanocrystalline metals with grain sizes below ~15 nm, grain boundary sliding and grain rotation control plasticity, with STZ-like events occurring within the intergranular region [34]. Such a mechanism can explain why nanocrystalline metals have pressure sensitive strengths (STZs are harder to operate under compression [35]) and why dopants which fully segregate to grain boundary sites can increase mechanical strength (by reducing grain boundary energy [23]). Our results presented here show that the properties of the crystalline lattice are important as well, with the Young's modulus of the overall material also influencing how difficult it is to induce plastic deformation. In addition to grain boundary deformation, the grain interior must also change shape and be plastically strained to maintain compatibility between grains. Figures 1b and c show this clearly, with significant nonzero strains found in the grain interiors. A stiffer lattice will make it more difficult for two grains to deform and rotate past each other, raising the yield strength.

In summary, MD simulations were used to study solid solution effects on the strength of nanocrystalline Cu-based alloys. For a variety of alloy chemistries, we find that yield strength is linearly related to the Young's modulus of the sample. This observation can provide a roadmap for the creation of extremely strong nanocrystalline alloys with grain sizes below ~15 nm, indicating that solute atoms which quickly stiffen the lattice are best and that changes to lattice spacing are largely inconsequential. The connection between strength and elastic modulus is reminiscent of metallic glass behavior, and provides yet another piece of evidence connecting nanocrystalline and amorphous metal deformation physics.

This work was supported by the U.S. Army Research Office, through Grant W911NF-12-1-0511. The author also thanks Prof. R. Birringer (Univ. Saarlandes) for helpful discussions about the mechanical properties of nanocrystalline alloys.

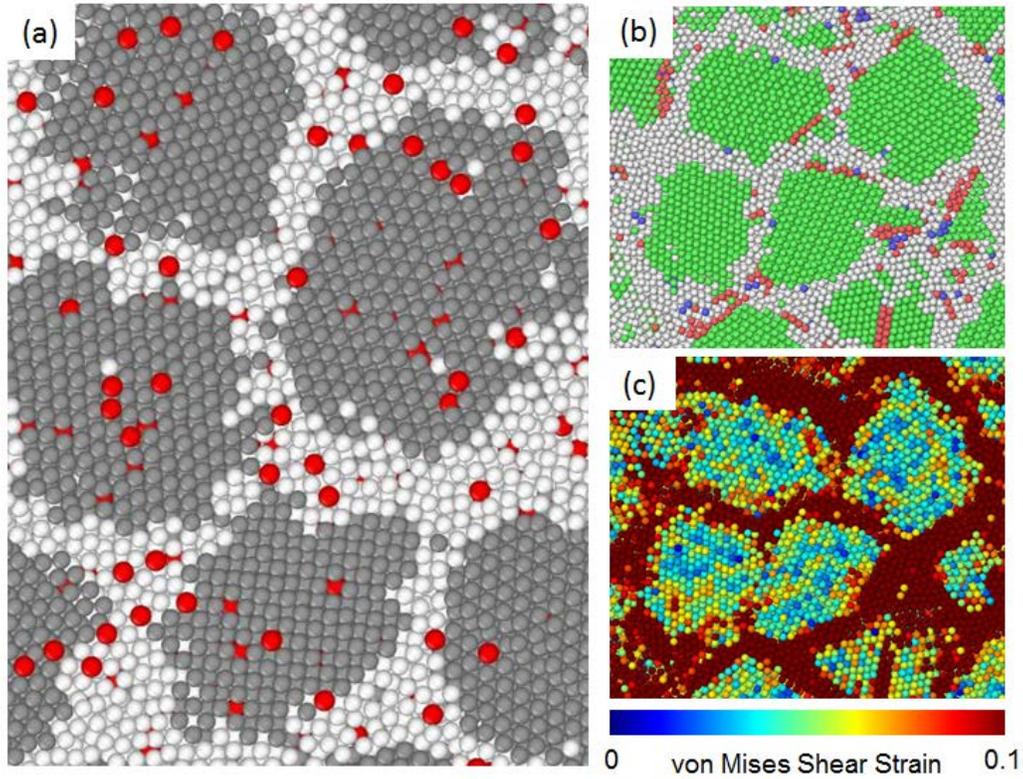

**Figure 1. (a) Atomic configuration of a Cu-4 at.% Pb alloy, showing a random solid solution. Red atoms denote Pb, while grain boundary and grain interior Cu atoms are colored white and grey, respectively. A deformed pure Cu sample is shown in (b) and (c) with the atoms colored according to CNA and von Mises shear strain, respectively.**

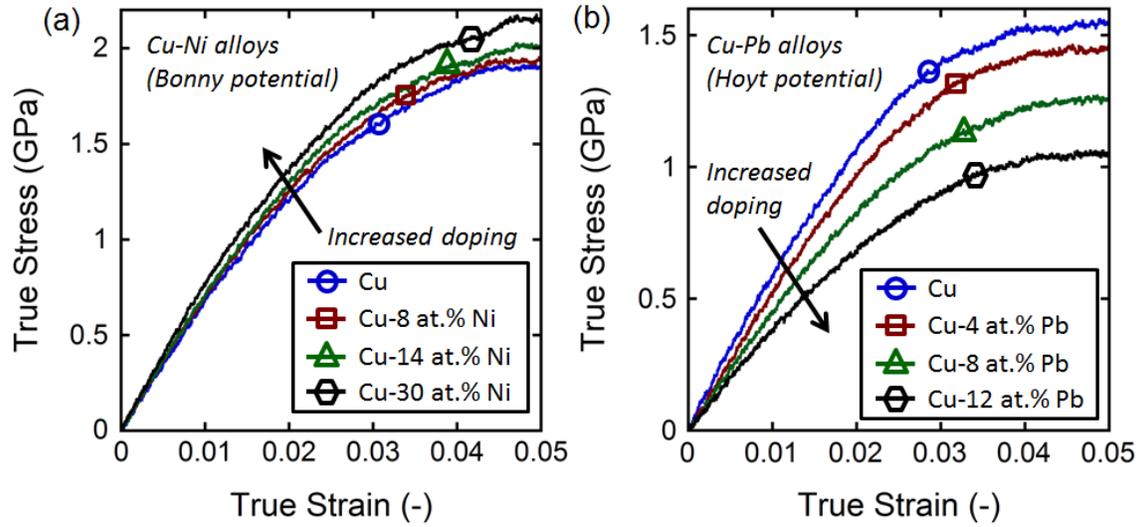

**Figure 2.** Tensile stress-strain curves for (a) Cu-Ni and (b) Cu-Pb alloys. While the addition of Ni strengthens nanocrystalline Cu, alloying with Pb weakens the material significantly.

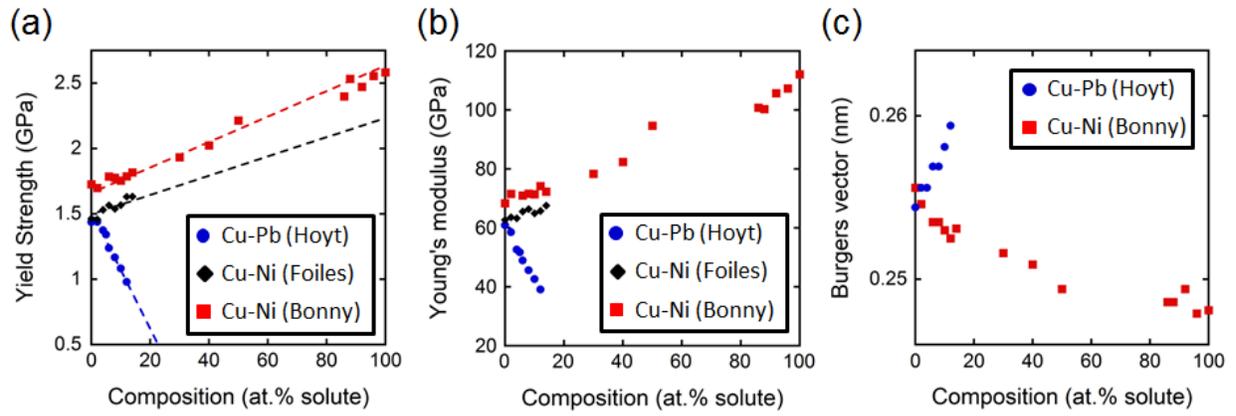

**Figure 3.** Compositional dependence of (a) yield strength, (b) Young's modulus, and (c) Burgers vector for all of the samples simulated in this study. Trends in strength mimic the observed changes in elastic properties of the system and are well-described by Eq. 2, shown as dotted lines in (a).

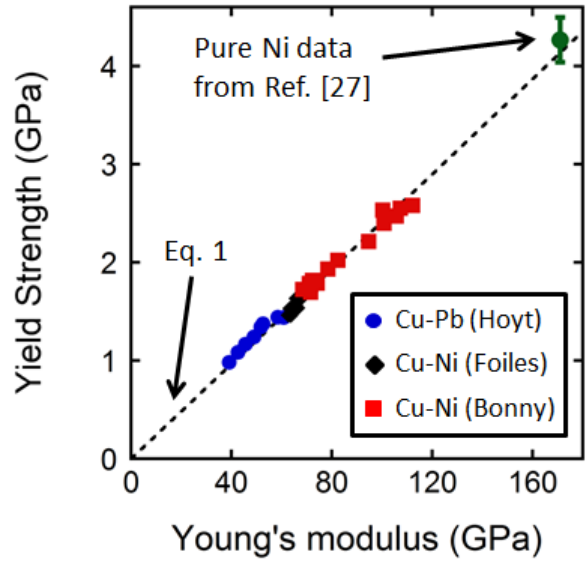

**Figure 4. Yield strength versus Young's modulus, showing a linear relationship following Eq. 1.**